%% file: MI.tex
\def\DIRvalue{Minahan}
\def\IDvalue{MI}
\def\titlevalue{Matrix models for 5d super Yang-Mills}
\def\authorvalue{Joseph A. Minahan}
\def\shortauthorvalue{\authorvalue}
\def\addressvalue{Department of Physics and Astronomy,
     Uppsala university,\\
     Box 516,
     SE-75120 Uppsala,
     Sweden\\
  \tt joseph.minahan@physics.uu.se}
\def\abstractvalue{ In this contribution to the review on localization in gauge theories we investigate the matrix models derived from localizing $\mathcal{N}=1$ super Yang-Mills on $S^5$.  We consider the large-$N$ limit and attempt to solve the matrix model by a saddle-point approximation.  In general it is not possible to find an analytic solution, but at the weak and  the strong limits of the  't Hooft coupling there are  dramatic simplifications that allows us to extract most of the interesting information.  At weak coupling we show that the matrix model is close to the Gaussian matrix model and that the free-energy scales as $N^2$.  At strong coupling we  show that if the theory contains one adjoint hypermultiplet then the free-energy scales as $N^3$.  We also find the expectation value of a supersymmetric Wilson loop that wraps the equator.  
 We demonstrate how to extract the effective couplings and reproduce results of Seiberg.  Finally, we compare to  results for the six-dimensional $(2,0)$ theory  derived  using the AdS/CFT correspondence.  We show that by choosing the hypermultiplet mass such that the supersymmetry is enhanced to $\mathcal{N}=2$,  the Wilson loop result matches the analogous calculation using AdS/CFT.  The free-energies differ by a rational fraction.
}
\def\preprintvalue{UUITP-15/16}

\ifx\ifLONG\undefined
\documentclass[11pt]{article}
\input{header.tex}

\input{MIdef.tex}
\newcommand{\Tr}{\textrm{Tr}}

\begin{document}
\thispagestyle{empty}
\documentheader
\else \chapterheader \fi

\numberwithin{equation}{section}

\newcommand{\sfrac}[2]{\mbox{$\frac{#1}{#2}$}}
\newcommand{\ms}{\!-\!}
\newcommand{\ps}{\!+\!}

\section{Introduction}

In this installment of the review on localization  we analyze the matrix models that result from localizing five-dimensional $\mathcal{N}=1$ super Yang-Mills on a five-sphere of radius $r$.  In five dimensions the supermultiplets have one vector multiplet and some hypermultiplets.  In this generic case there are a total of eight supersymmetries.  The most interesting case for us is when there is one hypermultiplet in the adjoint representation with a particular mass, $M$.  We will refer to such theories as $\mathcal{N}=1^*$ theories.  When  $M=i/(2r)$
 the supersymmetry is enhanced to $\mathcal{N}=2$, with 16 total supersymmetries.  This is the maximal amount of supersymmetry in five dimensions (without gravity), so we will refer to this as maximally supersymmetric Yang-Mills (MSYM).

The reason that the $\mathcal{N}=2$ enhancement is interesting is that the mysterious six-dimensional  $(2,0)$ superconformal field theory when compactified on a circle reduces to MSYM, with the radius related to the Yang-Mills coupling by
 \begin{equation}\label{MIR6gym}
 R_6=\frac{g_{YM}^2}{8\pi^2}\,.
 \end{equation}
 This relation follows from identifying the Kaluza-Klein modes of the $(2,0)$ theory with the instanton particles in the 5d MSYM \cite{MIWitten:1995ex}.   
These $(2,0)$ theories are difficult to study because they have no free parameters and no Lagrangian description, and thus no perturbative prescription.  However, they do have an $AdS_7\times S^4$ dual, so certain strong coupling data can be extracted using supergravity.  The hope then is that one can use the localization results from the MSYM to say something about the $(2,0)$ theory.   For example, one can now say much about the supersymmetric indices of the (2,0) theories using MSYM (see  \volcite{KL}).

One thing to keep in mind about this discussion is that the MSYM is non-renormalizable and hence requires a UV completion.  The $(2,0)$ theory on the circle is believed to be a consistent UV completion \footnote{It had been proposed that MSYM could be used to actually define the $(2,0)$ theories \cite{MIDouglas:2010iu,MILambert:2010iw,MIBolognesi:2011nh},  and while not renormalizable, might actually be finite \cite{MIDouglas:2010iu}.  However, an explicit calculation shows that the four-point amplitude is UV divergent at six loops \cite{MIBern:2012di} and hence requires a UV completion. For a possible way around this see \cite{Papageorgakis:2014dma}.}.  The observables we compute using localization are however finite because of the supersymmetry and would be expected to match with the same observables in the UV complete theory.

Localization results in a complicated matrix model that is not analytically solvable in general.  However, we will show that in the large-$N$ limit at strong coupling
 the analysis of the matrix model simplifies dramatically.  One of the main results is that free-energy scales as $N^3$ \cite{MIKim:2012av,MIKallen:2012zn} with a coefficient that depends on $M$ \cite{MIKallen:2012zn}.  The supergravity analysis of the $(2,0)$ theory also exhibits  $N^3$ behavior for the free-energy \cite{MIKlebanov:1996un,MIHenningson:1998gx}, suggesting that the degrees of freedom are more than for a weakly coupled gauge theory, where one finds $N^2$ scaling in the free-energy.  However, at the MSYM point, the coefficient in front of the $N^3$ term differs by a factor of $4/5$ with the $N^3$ term in the supergravity calculation.  This remains an unresolved problem.
 
Nevertheless one can study another supersymmetric observable, the expectation value of a Wilson loop along the equator.  Here one finds a match at the MSYM point with a parallel computation done using supergravity \cite{MIMinahan:2013jwa}.

The rest of the paper is organized as follows.  In section \ref{MIS-general-matrix} we give some  details of the matrix model resulting from localization of 5d SYM and study  limits at large volume or large hypermultiplet mass, reproducing results in \cite{MISeiberg:1996bd,MIIntriligator:1997pq} for the effective couplings.  In section \ref{MIadjoint-5d-YM} we consider  the large $N$ behavior of the  $\mathcal{N}=1^*$ theories. 
We calculate the free energy  and the expectation value 
  of a supersymmetric  Wilson loop in the weak and strong coupling limits.  We also generalize these results to  $\mathbb{Z}_k$ quiver theories.   In section \ref{MIs-sugra} we compute 
   the free energy and the Wilson loop expectation value starting from the supergravity dual of the $(2,0)$ theory on $S^5\times R$, and then compactifying the $R$ to an $S^1$ and identifying the radius as in (\ref{MIR6gym}).  We show that there is a mismatch with the free-energy result from section \ref{MIadjoint-5d-YM} by a factor of $4/5$, but the Wilson loop results agree.
   In section \ref{MIsummary} we give a brief summary discuss some open problems.

 \section{Matrix model for $\mathcal{N}=1$ $5d$ Yang-Mills with matter}
 \label{MIS-general-matrix}

The perturbative partition function was derived in  \cite{MIKallen:2012va} for massless hypermultiplets and in
 \cite{MIKim:2012ava} for  MSYM.  Its derivation is given in \volcite{QZ}. In this section we show how the results of the effective couplings in \cite{MISeiberg:1996bd,MIIntriligator:1997pq} can be extracted from the resulting matrix model. 
 
We consider a theory with a semi-simple compact gauge group $G$. This has
an $\mathcal{N}=1$ vector multiplet and  $\mathcal{N}=1$ massless hypermultiplets in representation $\rm{R_i}$ with splittings into half-multiplets when $R_i$ is complex.  
 The partition function of this gauge theory on $S^5$ is then given by
     \begin{equation}\label{MIpathint}
Z=\int\limits_{\rm Cartan} [d\phi]~e^{-  \frac{ 8 \pi^3 r}{g_{YM}^2}  \text{Tr}(\phi^2)-\frac{\pi k}{3}\text{Tr}(\phi^3)}  Z_{\rm 1-loop}^{\rm vect} (\phi)    \prod_{i}Z_{\rm 1-loop}^{i} (\phi) + \mathcal{O} (e^{-\frac{16 \pi^3 r}{g_{YM}^2}})\,,
\end{equation}
 where the one-loop contributions are given by the  infinite products
\begin{equation}
 Z_{\rm 1-loop}^{\rm vect} (\phi) =\prod\limits_\beta\prod\limits_{t \neq 0}\left( t - \langle \beta,i\phi\rangle   \right)^{(1+\frac{3}{2}t+\frac{1}{2}t^2)}\,,\label{MIvect1-loop-beg}
\end{equation}
 and
\begin{equation}\label{MImain-form-det}
Z_{\rm 1-loop}^{i} (\phi) = \prod\limits_{\mu\in R_i} \prod\limits_{t}\left( t - \langle i\phi , \mu\rangle +\frac{3}{2} \right)^{-(1+\frac{3}{2}t+\frac{1}{2}t^2)}\,,
\end{equation}
with $\beta$  the roots and  $\mu$   the weights in  $\rm{R_i}$. 

The path integral in (\ref{MIpathint})  has a contribution from a Chern-Simons term with level $k$.
We have also absorbed the radius $r$
  into the integration variable $\phi = -i r\sigma$.   As in the 4D case  \cite{MIPestun:2007rz}, we must Wick rotate and integrate over real $\phi$ in order  to have a well-defined path integral.

The infinite products that appear in (\ref{MIvect1-loop-beg}) and (\ref{MImain-form-det}) are divergent and need to be regularized.   Each one-loop contribution has the form
  \begin{equation}
{\cal P}=  x \prod_{t=1}^\infty  \left ( t+ x \right)^{(1+\frac{3}{2}t+\frac{1}{2}t^2)} \left ( t- x \right)^{(1-\frac{3}{2}t+\frac{1}{2}t^2)}\,,\label{MIinfin-prod}
  \end{equation}
 whose log can be written as
\begin{equation}
\log {\cal P} = \sum_{t=1}^\infty \left ( 3x - \frac{x^2}{2} \right ) + {\rm convergent~ part}\,.
\end{equation}
Therefore, the infinite product can be regulated by replacing it with the triple sine function \cite{MIkurokawa}
 \begin{equation}
S_3(x) = 2\pi e^{-\zeta'(-2)} x e^{\frac{x^2}{4} - \frac{3}{2}x} \prod_{t=1}^\infty \Big ( \left ( 1+ \frac{x}{t} \right)^{(1+\frac{3}{2}t+\frac{1}{2}t^2)} \left ( 1- \frac{x}{t} \right)^{(1-\frac{3}{2}t+\frac{1}{2}t^2)}
 e^{\frac{x^2}{2} - 3x} \Big )\,,\label{MIdef-triple-sine}
\end{equation}
   As an alternative we can regularize the divergence by introducing  a UV cut-off that stops the mode expansion at  $n_0 = \pi \lambda r$,  
and leaving the log of the one-loop determinants to be
\begin{eqnarray}
\log ( Z_{\rm 1-loop}^{\rm vect} (\phi) \prod_i Z_{\rm 1-loop}^{i} (\phi) ) &=&  -\frac{\pi\lambda r}{2}
 \sum\limits_{\beta} (\langle \beta,i\phi\rangle )^2 +   \frac{\pi\lambda r}{2}
 \sum\limits_{\mu\in {\rm R}_i} ( \langle i\phi , \mu\rangle)^2 +   {\rm convergent~ part}\nonumber \\
& = &\pi \lambda r \left (C_2 ({\rm adj}) - \sum_iC_2(\rm{R}_i) \right ) \text{Tr}(\phi^2 ) + {\rm convergent~ part}\,,\nonumber\\
\end{eqnarray}
 where  
 $\Tr (T_A T_B) = C_2(R) \delta_{AB}$ and 
  $\sum\limits_{\mu\in {\rm R}_i} ( \langle \phi , \mu\rangle)^2 = 2 C_2({\rm R}_i) \Tr (\phi^2)$. 
  The linear 
   piece cancels since the gauge group is  semi-simple. Hence,  the divergent 
    piece is proportional to $\text{Tr}(\phi^2)$ and  can be absorbed into an effective coupling given by
\begin{equation}
 \frac{1}{g_{eff}^2} = \frac{1}{g_{YM}^2} - \frac{\lambda}{ 8\pi^2} \left (C_2 ({\rm adj}) - \sum_iC_2(\rm{R}_i) \right)\,.  
 \end{equation}
This  renormalization of the coupling agrees 
  with the  flat space results in \cite{MINekrasov:1996cz,MIFlacke:2003ac}. 
 
 The convergent part of  (\ref{MIinfin-prod}) can be replaced by $S_3(x) e^{-\frac{x^2}{4} + \frac{3}{2} x}$
   up to  $x$-independent (and hence irrelevant) constants. 
 The extra exponential factor leads to a  further finite shift in the coupling constant.  Notice that the UV divergence  cancels if there is only one hypermultiplet and it sits in the adjoint representation.  

Using the regularized determinants, 
  we can rewrite the matrix model  in terms of triple sine  functions $S_3(x)$
 \begin{eqnarray}
 Z=  \int d\phi e^{-\frac{ 8 \pi^3 r}{g_{YM}^2}  \Tr (\phi^2)-\frac{\pi k}{3}\text{Tr}(\phi^3)} {\rm det}_{Ad} \Big ( S_3(i\phi)  \Big )\
   \prod_i{\rm det}^{-1}_{{\rm R}_i} \Big ( S_3 \left (i\phi + \frac{3}{2} \right  ) \Big )\,,
 \end{eqnarray}
  where from now on we assume that $g_{YM}$ is the renormalized coupling.  
 The   triple sine function $S_3(x)$ has the symmetry property
  \begin{equation}
   S_3( - x) = S_3(x+3)~,~~~~~~S_3\left (x+\frac{3}{2} \right ) = S_3\left (- x+\frac{3}{2} \right )\,.
  \end{equation}
The weights are mapped from   $\mu$ to $-\mu$ when exchanging representation $\rm{R}$ with $\bar {\rm{R}}$. Hence,  the one-loop contribution of a massless hypermultiplet has the property
  \begin{equation}
    {\rm det}_{R} \Big ( S_3 \left (i\phi + \frac{3}{2} \right  )   \Big )=  {\rm det}_{R} \Big ( S_3 \left (-i\phi + \frac{3}{2} \right  ) \Big )
    = {\rm det}_{\bar{R}} \Big ( S_3 \left (i\phi + \frac{3}{2} \right  ) \Big )\,.
  \end{equation}
   and the representations $\rm{R}$ and $\rm{\bar {R}}$ are automatically symmetrized in the determinants.
     
 Hypermultiplet masses can be turned on  
    by using an auxiliary $U(1)$ vector multiplet. 
One takes a  
      $G \times U(1)$ matrix model, but excludes the integration over the $U(1)$ direction.  Thus the contribution of massive 
      hypermultiplets is given by
        \begin{equation}
 Z=  \int d\phi\, e^{-\frac{ 8 \pi^3 r}{g_{YM}^2} \Tr (\phi^2)- \frac{\pi k}{3} \Tr (\phi^3)} {\rm det}_{Ad} \Big ( S_3(i\phi)  \Big )\
 \prod_i  {\rm det}^{-1}_{{\rm R}_i} \Big ( S_3 \left (i\phi +i m_i + \frac{3}{2} \right  ) \Big )\,,
   \end{equation}
where $m_i$ are dimensionless parameters related to the actual hypermultiplet masses by  $m_i= rM_i$.     Using  the triple sine's symmetry  we find the relation
   \begin{equation}
    {\rm det}_{{\rm R}_i} \Big ( S_3 \left (i\phi + i m_i + \frac{3}{2} \right  )   \Big ) =   
     {\rm det}_{\bar{{\rm R}_i}} \Big ( S_3 \left (i\phi - i m_i + \frac{3}{2} \right  )   \Big )\,. 
    \end{equation}
     Hence, the partition function with  massive hypermultiplets  can be  written as  
     \begin{eqnarray}\label{MImm1}
 && \int d\phi\, e^{-\frac{ 8 \pi^3 r}{g_{YM}^2} \Tr (\phi^2)-\frac{\pi k}{3} \Tr(\phi^3)} {\rm det}_{Ad} \Big ( S_3(i\phi)  \Big )\  \nonumber \\
 &&\qquad\qquad  \times\prod_i {\rm det}^{-1/2}_{{\rm R}_i} \Big ( S_3 \left (i\phi + i m_i +  \frac{3}{2} \right ) \Big ) {\rm det}^{-1/2}_{\bar{{\rm R}_i}}\Big ( S_3 \left (i\phi - i m_i +  \frac{3}{2} \right  )  \Big )\,. \label{MIfull-matrix-model}
   \end{eqnarray}
  
Let us consider (\ref{MImm1})  in the large volume limit by taking $r\to\infty$.
We can write (\ref{MImm1}) in the  form
 \begin{equation}
   \int d\phi~  e^{-{\cal F}}\,,
 \end{equation}
 where 
 \begin{equation}
 {\cal F} = \frac{ 8 \pi^3 r}{g_{YM}^2} \Tr (\phi^2)+\frac{\pi k}{3}\Tr(\phi^3) - \sum_\beta \log S_3( \langle i\phi, \beta\rangle ) + \sum_i\sum\limits_{\mu_i} \log S_3 \Big (\langle i\phi, \mu_i \rangle + i\,m_i + \frac{3}{2} \Big )\,. \label{MIfull-F-prep}\nonumber\\
 \end{equation}
  We then restore the $r$   dependence by the rescaling
  $\phi \rightarrow r\phi$ and $m \rightarrow rm$. Using 
 the asymptotic expansion  for $| \text{Im} z| \rightarrow \infty$ and $0\leq \text{Re} z < 3$ 
  \begin{equation}
 \log S_3(z) \sim - \text{sgn} ( \text{Im} z) \pi i \left ( \frac{1}{6} z^3 - \frac{3}{4}z^2 + z + ...   \right )\,,
 \end{equation}
 we obtain the expression
 \begin{equation}\label{MIprepot}
\frac{1}{2\pi r^3}  {\cal F} =  \frac{ 4  \pi^2}{g_{YM}^2} \Tr (\phi^2)+\frac{ k}{6}\Tr(\phi^3) + 
\frac{1}{12} \Big ( \sum_\beta |\langle \phi, \beta\rangle |^3 - \sum_i\sum\limits_{\mu_i}  | \langle \phi, \mu_i \rangle + m_i|^3 \Big )+ O(r^{-2})\,.
 \end{equation}
  Up to  a constant which we have absorbed into the definition  of the coupling, (\ref{MIprepot}) matches the   quantum prepotential
   in the flat-space limit  \cite{MIIntriligator:1997pq}.   The normalization  of the quadratic term is fixed either by a direct one-loop calculation in  flat space  \cite{MIFlacke:2003ac} or by matching the superpotential in 5d with the corresponding one in 4D  \cite{MINekrasov:1996cz}.

The matrix model is well-defined if
  ${\cal F}$ is a convex positive  function in the large $\phi$ limit.
  In this limit $\mathcal{F}$ takes the asymptotic form
    \begin{eqnarray}\label{MIFF}
    {\cal F} &= & \frac{ 8 \pi^3 r}{g_{YM}^2} \Tr (\phi^2)+\frac{\pi k}{3}\Tr(\phi^3) + 
\frac{\pi}{6} \Big ( \sum_\beta |\langle \phi, \beta\rangle |^3 - \sum_i\sum\limits_{\mu_i} | \langle \phi, \mu_i \rangle|^3 \Big )\nonumber \\
&& - \pi \sum_\beta |\langle \phi, \beta\rangle |  -\frac{\pi}{2} \sum_i\sum_{\mu_i}\left(m_i \, \text{sgn} (  \langle \phi, \mu_i \rangle)  (\langle \phi, \mu_i \rangle)^2+
 ( m_i^2 + \frac{1}{4}  )  |\langle \phi, \mu_i \rangle | \right)+ \cdots\,,\nonumber\\
    \end{eqnarray}
    where the ellipsis stands for  terms suppressed at large   $\phi$.  Analyzing the  convexity of (\ref{MIFF}) it is clear that 
     the cubic terms dominate.  Hence, the analysis is identical to that in \cite{MIIntriligator:1997pq} and  the same conditions apply.  In  special cases  the cubic terms  cancel each other, for example 
     in the case of single adjoint hypermutiplet 
       \cite{MIKallen:2012zn} or for the superconformal $USp(2N)$ theory described in
   \cite{MIJafferis:2012iv}.

Suppose we now take the hypermultiplet masses to infinity.  For large $|m_i|$ the leading terms in (\ref{MIfull-F-prep}) are
  \begin{eqnarray}\label{MIF:def}
  {\cal F} &= &\frac{ 8 \pi^3 r}{g_{YM}^2} \Tr (\phi^2)+\frac{\pi k}{3}\Tr(\phi^3) - \sum_\beta \log S_3( \langle i\phi, \beta\rangle ) \nonumber\\
 &&\qquad\qquad- \sum_i\text{sgn} (m_i) \frac{\pi}{2} 
   \sum\limits_{\mu_i} \left ( \frac{1}{3} ( \langle \phi, \mu_i \rangle )^3 + m (\langle \phi, \mu_i \rangle)^2 \right )\,.\nonumber
 \end{eqnarray}
The two last terms in (\ref{MIF:def}) can be absorbed by a redefinition of $k$ and $g_{YM}$.
   To see this,  note that
    \begin{equation}\label{MItrace3}
     \text{Tr} (T_A T_B T_C + T_A T_C T_B) = C_3( {\rm R} ) d_{ABC}\,,
    \end{equation}
  where 
  \begin{equation}\label{MIC3}
  \sum\limits_\mu   ( \langle \phi, \mu \rangle )^3 = C_3 ( {\rm R} ) \text{Tr} (\phi^3)\,. 
  \end{equation}
 The coefficient $C_3$ satisfies $C_3( \bar{{\rm R}} ) = - C_3( {\rm R} )$, hence it is zero for real representations.  For the lower complex representations  in $SU(N)$ it is  $1$ for the fundamental,  $N-4$ for the antisymmetric, and $N+4$ for the symmetric representations.  
    Hence, from (\ref{MItrace3}) and (\ref{MIC3}) we get 
    \begin{equation}
     k_{eff} = k  - \sum_i\text{sgn}(m_i) \frac{C_3 ( {\rm R}_i )}{2}\,,
    \end{equation}
   the same  result in  \cite{MISeiberg:1996bd,MIIntriligator:1997pq}.
A similar analysis of the quadratic terms gives  
\begin{equation}
 \frac{r}{g_{eff}^2} = \frac{r}{g_{YM}^2} - \sum_i\frac{|m_i|}{ 8\pi^2} C_{2}({\rm R}_i)\,.
\end{equation}

\section{$\mathcal{N}=1^*$ $5d$ super Yang-Mills}
\label{MIadjoint-5d-YM}

We now turn to $\mathcal{N}=1^*$  super Yang-Mills, where there is a single adjoint hypermultiplet with mass parameter $m$.   We further assume that the gauge group is $SU(N)$.
 
To analyze the resulting matrix model (\ref{MIfull-matrix-model}) we  rewrite the triple sine function $S_3(z)$  in (\ref{MIdef-triple-sine})
 as 
 \begin{eqnarray}
  S_3(z) = 2 e^{-\zeta'(-2)} \sin (\pi z) ~e^{\frac{1}{2} f(z)} ~e^{\frac{3}{2} l(z)}~,
 \end{eqnarray}
 where $l(z)$  and $f(z)$ are given by  \cite{MIJafferis:2010un,MIKallen:2012cs}
 \begin{eqnarray}
l(z)&=&-z\log\left(1-e^{2\pi i z}\right)+\frac{i}{2}\left(\pi z^2+\frac{1}{\pi}\mathrm{Li}_2(e^{2\pi i z})\right)
-\frac{i \pi}{12}
\label{MIl:function}
\\
f(z)&=&\frac{i\pi z^3}{3}+z^2 \log\left(1-e^{-2\pi i z}\right)+\frac{i z}{\pi}\mathrm{Li}_{2}\left(e^{-2\pi i z}\right)
+\frac{1}{2\pi^2}\mathrm{Li}_{3}\left(e^{-2\pi i z}\right)-\frac{\zeta(3)}{2\pi^2}\,.
\label{MIf:function}
\end{eqnarray}
While these functions are rather ugly, their derivatives have the much nicer form
    \begin{equation}
\frac{d f(z)}{dz}=\pi z^2 \cot(\pi z)\, ;\qquad \frac{d l(z)}{dz}=-\pi z \cot(\pi z) \,;
\label{MIf:l:derivative}
\end{equation}
The matrix model path integral (\ref{MIfull-matrix-model}) can then be rewritten as 
\begin{eqnarray}
\nonumber
Z& =&\int\limits_{Cartan}\left[d\phi\right]e^{-\frac{ 8 \pi^3 r}{g_{YM}^2}\mathrm{Tr}(\phi^2)}\prod_{\beta}
(\sin(\pi\langle\beta, i\phi\rangle) e^{-
\frac{1}{4}l(\frac{1}{2}-i m -\langle\beta, i\phi\rangle)-
\frac{1}{4}l(\frac{1}{2}-i m +\langle\beta, i\phi\rangle)}
\\
&&\qquad\qquad\qquad\qquad\qquad\times  e^{\frac{1}{2}f(\langle\beta, i\phi\rangle)-
\frac{1}{4}f(\frac{1}{2}-i m -\langle\beta, i\phi\rangle)-
\frac{1}{4}f(\frac{1}{2}-i m +\langle\beta, i\phi\rangle)} + \cdots~,
\label{MImain-matrix123}
\end{eqnarray}
up to instanton terms, where  we have  dropped the Chern-Simons term.  From now on we assume that the gauge group is $U(N)$.
Defining the t' Hooft coupling constant to be
$$\lambda=\frac{g_{YM}^2 N}{r}\,,$$
and taking the large  $N$ limit for fixed $\lambda$, all instanton contributions are suppressed.  We can then re-express (\ref{MImain-matrix123}) as the integral over the  eigenvalues $\phi_i$
\begin{eqnarray}
\nonumber
& &Z \sim  \int \prod_{i=1}^{N}d\phi_{i}\exp\left(-\frac{ 8 \pi^3 N}{\lambda}\sum\limits_{i}\phi_{i}^2+
\sum\limits_{j\neq i}\sum\limits_{i}\biggl[\log\left[\sinh(\pi(\phi_i-\phi_j))\right]
\biggr.\right.
\\
\nonumber
& & \qquad
-\frac{1}{4}l\left(\frac{1}{2}-i m+i(\phi_i-\phi_j)\right)
 -\frac{1}{4}l\left(\frac{1}{2}-i m-i(\phi_i-\phi_j)\right)+ \frac{1}{2}f(i(\phi_i-\phi_j))-
\\
& &\qquad\left.\left.
-\frac{1}{4}f\left(\frac{1}{2}-i m+i(\phi_i-\phi_j)\right)-
\frac{1}{4}f\left(\frac{1}{2}-i m-i(\phi_i-\phi_j)\right)\right]\right)~. 
\label{MIpartition:function}
\end{eqnarray}

In the large  $N$ limit the partition function in (\ref{MIpartition:function}) is dominated by the saddle point.  Using the derivatives in (\ref{MIf:l:derivative}),   the $\phi_i$ at the saddle point  satisfy 
\begin{eqnarray}
\nonumber
\frac{ 16 \pi^3 N}{\lambda}\phi_i&=& \pi \sum\limits_{j\neq i}\Bigg[\left(2- (\phi_i-\phi_j)^2\right)\coth(\pi(\phi_i-\phi_j))
\\
&&\qquad\qquad+\frac12\left(\frac{1}{4}+(\phi_i-\phi_j-m)^2\right)\tanh(\pi(\phi_i-\phi_j- m))\nonumber\\
&&\qquad\qquad+\frac12\left(\frac{1}{4}+(\phi_i-\phi_j+ m)^2\right)\tanh(\pi(\phi_i-\phi_j+ m))\Bigg]\,.
\label{MIeom}
\end{eqnarray}
In general this equation is not solvable, but it simplifies a lot  both  at weak  ($\lambda\ll1$) and strong  $(\lambda\gg1$) coupling.

For weak coupling, the contribution from the hypermultiplet plays no role and (\ref{MIeom}) reduces to
\begin{equation}
\frac{16 \pi^3 N}{\lambda}\phi_i\approx 2\sum_{j\ne i}\frac{1}{\phi_i-\phi_j}\,.
\end{equation}
This is the same equation one finds for a Gaussian matrix model and in the large-$N$ limit its  solution has the  Wigner distribution
\begin{equation}\label{MIrho-weak}
\rho(\phi)\equiv\frac{1}{N}\frac{dn}{d\phi}=\frac{2}{\pi\phi_0^2}\sqrt{\phi_0^2-\phi^2}\,,\qquad\ \phi_0=\sqrt{\frac{\lambda}{4 \pi^3}}\,,
\end{equation}
where the eigenvalue density is normalized to
\begin{equation}
\int \rho(\phi)d\phi=1~.
\end{equation}
The free energy then has the typical weak coupling form
\begin{equation}
F=-\log Z\approx -N^2\log \sqrt{\lambda}\,.
\end{equation}

\begin{figure}
\begin{center}
  \includegraphics[width=53mm,angle=0,scale=1.3]{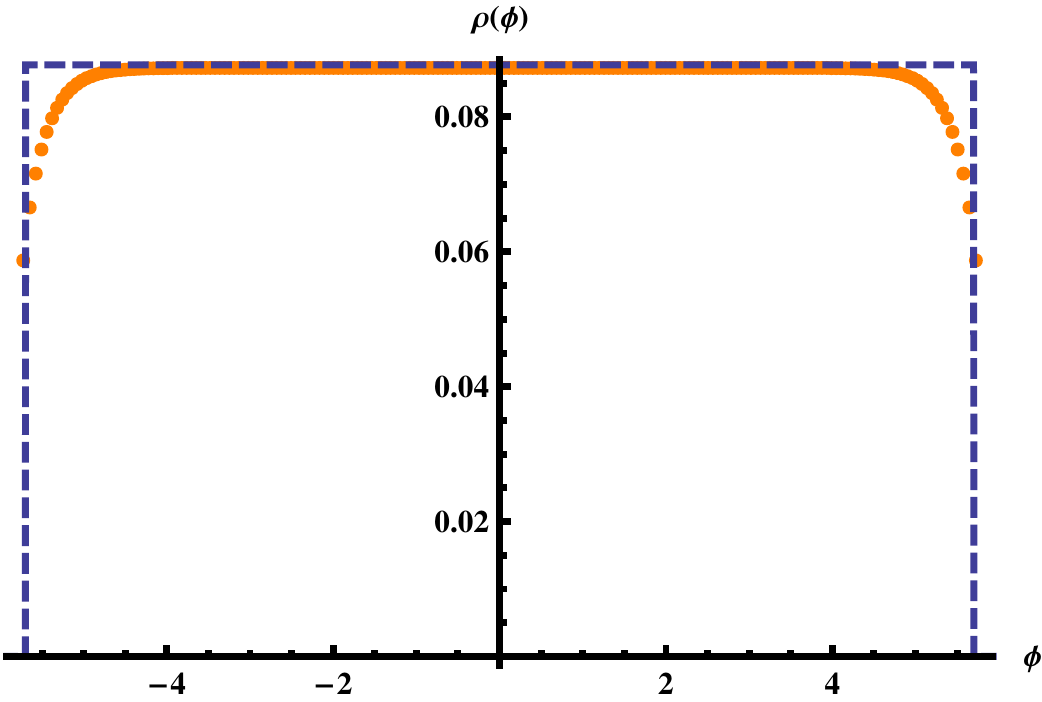}\hspace{5mm}
  \includegraphics[width=50mm,angle=0,scale=1.37]{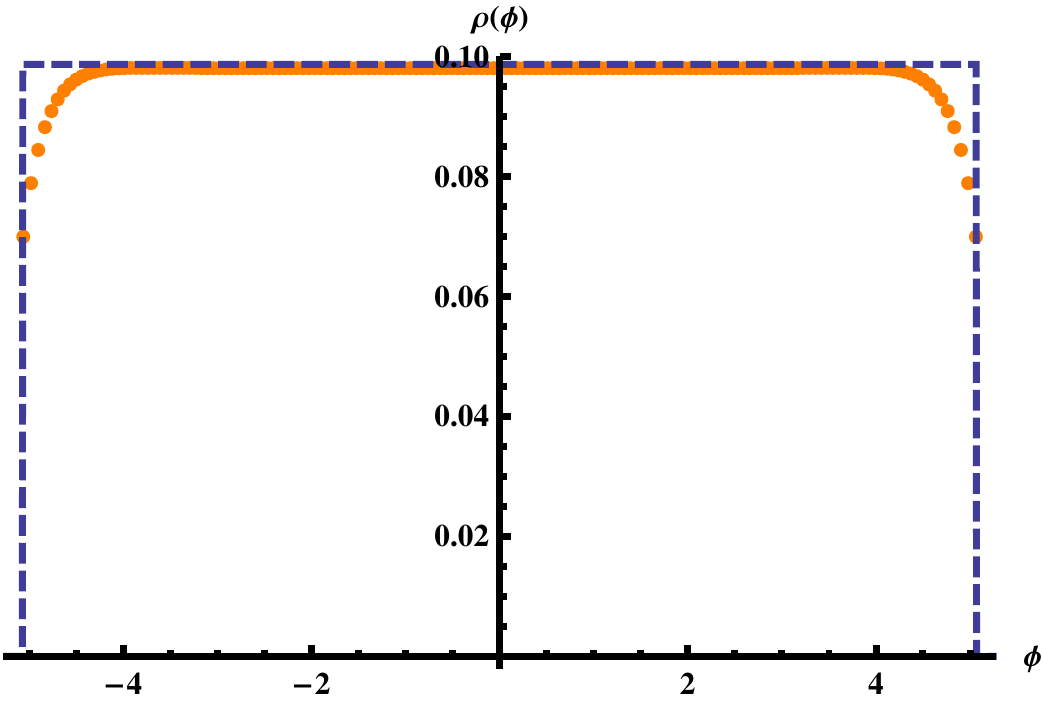}
\end{center}
\caption{Numerical results for the eigenvalue density  $\rho(\phi)$ for (left) $m=0$, $N=200$, 
$\lambda=400$ and (right) $m=\frac{1}{2}$, $N=160$, 
$\lambda=320$. The dashed blue lines are the  strong coupling solutions in (\ref{MIrho-strong}).}
\label{MIdensity:pic}
\end{figure}

At strong coupling and with $|m| \ll\lambda$ we can simplify (\ref{MIeom}) by making the ansatz \mbox{$|\phi_i-\phi_j|\gg1$}.  In general this is not the case for every pair  $(i,j)$, but it will be true for most of them.  The saddle point equation then simplifies to
\begin{equation}
\frac{16 \pi^3 N}{\lambda}\phi_i=\pi
\left(\frac{9}{4}+ m^2\right)\sum\limits_{j\neq i}\mathrm{sgn}(\phi_i-\phi_j)\,.
\label{MIeom:strong}
\end{equation}
If we assume  the $\phi_i$  are ordered, we get 
\begin{equation}
\phi_i=\frac{\left(9+4m^2\right)\lambda}{64\pi^2 N}(2i-N)\,.
\label{MIsolution:strong}
\end{equation}
Hence the  eigenvalue density is constant over its support,
\begin{eqnarray}\label{MIrho-strong}
\rho(\phi)&=&\frac{32 \pi^2 }{\left(9+4 m^2\right)\lambda}\,,\qquad |\phi|\le\phi_m~,\qquad \phi_m=\frac{\left(9+4m^2\right)\lambda}{64\pi^2}\nonumber\\
&=&\qquad 0\qquad \qquad |\phi|>\phi_m\,.
\end{eqnarray}
In figure \ref{MIdensity:pic} we compare the approximation in (\ref{MIrho-strong}) with numerical results using the full kernel in (\ref{MIeom}).  As one can see the approximation is very good except with a slight deterioration at the end-points. 

Using the strong-coupling ansatz, the partition function in  (\ref{MImain-matrix123})  simplifies to
\begin{equation}
Z\sim\int\prod_{i} d\phi_{i} e^{-\frac{ 8\pi^3N}{\lambda}\sum\limits_{i}\phi_{i}^2+\frac{\pi}{2}
\left(\frac{9}{4}+m^2\right)\sum\limits_{j\neq i}\sum\limits_{i}|\phi_i-\phi_j|}\,.
\label{MIpartition:strong}
\end{equation}
Applying the saddle point solution (\ref{MIsolution:strong}),
we find for the free   energy
\begin{equation}
 F \equiv -\log Z\approx -\frac{g_{YM}^2 N^3}{ 96\pi r}\left(\frac{9}{4}+m^2\right)^{2}+\mbox{O}(N^2)
 \,.
 \label{MImatrix:free:energy}
\end{equation}
Thus we see that the free energy crosses over from the $N^2$ dependence expected in a weakly coupled gauge theory to an $N^3$ behavior  when going from weak to strong coupling.

The other interesting observable we can compute using localization is a supersymmetric Wilson loop \cite{MIKim:2012qf,MIAssel:2012nf,MIMinahan:2013jwa}.
Such loops in five-dimensional flat space were first considered in \cite{MIYoung:2011aa}.  
On $S^5$, the loop must run along an
 $S^1$ fiber  over a base $\mathbb{C}P^2$ in order to preserve some of the supersymmetries.  The supersymmetric Wilson loop is  given by
 \begin{equation}
\langle W\rangle =\frac{1}{N}\left\langle \Tr \left(\mbox{P}\exp(i\oint_{S^1}(A_M v^M))ds\right)\right\rangle\,.
\end{equation}
where we have written the bosonic fields $A_M$ in the 10D notation of \cite{MIPestun:2007rz,MIMinahan:2015jta}.  The vector $v^M$ is the 10-dimensional vector $v^M=\epsilon\Gamma^M\epsilon$ where the components $M=1,\dots5$ reduce to the Reeb vector on $S^5$ and $v^0=1$, where $A_0=\sigma$.  All other components of $v^M$ are zero.  Hence, along the localization locus in the zero instanton sector we have $A_Mv^M=\sigma$. After Wick rotation the Wilson loop then becomes
\begin{equation}
\langle W\rangle=\frac{1}{N}\langle \Tr\, e^{2\pi\phi}\rangle=\frac{1}{N}\left\langle \sum_i\, e^{2\pi\phi_i}\right\rangle\,.
\end{equation}
In the large $N$ limit the $\sum\limits_{i}e^{2\pi\phi_{i}}$ term has a negligible back-reaction on the 
saddle point solutions.
Thus in this limit the  Wilson loop  is well-approximated by the 
integral
\begin{equation}
\langle W\rangle = \int d\phi \rho(\phi)e^{2\pi\phi}\,.
\end{equation}

At weak coupling where the density has support only for $|\phi|\ll1$ we can approximate the integral as
\begin{equation}
\langle W\rangle \approx \int d\phi \,\rho(\phi)(1+2\pi^2\phi^2)=1+\frac{\lambda}{ 8 \pi}\approx \exp\left({\frac{\lambda}{8\pi}}\right).
\end{equation}
At strong coupling, where the density is approximately constant along its support, we find
\begin{equation}
\langle W\rangle \approx  \frac{ 32 \pi^2 }{\left(9+4m^2\right)\lambda}\int\limits_{-\phi_m}^{\phi_m}e^{2\pi\phi}d\phi\sim
\exp\left({\frac{\lambda}{ 8\pi}\left(\frac{9}{4}+m^2\right)}\right),	
\label{MIwilson:loop:matrix}
\end{equation}
where we have dropped the prefactor since it can be affected by our approximation and besides is  not particularly important for the rest of the discussion.    Interestingly, and unlike 4D, the argument in the exponent still has linear $\lambda$ dependence at strong coupling, with the coefficient changed by the factor  $\left(\frac{9}{4}+m^2\right)$ as compared to the  weak coupling result. 

The results for the free-energy and the Wilson loop can be generalized to a $\mathbb{Z}_k$ quiver of the $\mathcal{N}=1^*$ theory \cite{MIKallen:2012zn,MIMinahan:2013jwa}.  The quiver has an $SU(N)^k$ gauge group with  equal mass hypermultiplets  transforming in the bifundamental representations,
$(N,\overline{N},1,\dots 1)$, $(1,N,\overline{N},1,\dots )$, {\it etc.}   The eigenvalues of (\ref{MIeom})  divide into $k$ groups $\psi^{(r)}_i$, where $r=1,\dots,k$, $i=1,\dots N$,  resulting in the equations of motion 
\begin{eqnarray}
\frac{ 16\pi^3 N}{\lambda}\psi^{(r)}_i&=& \pi \Big[\sum\limits_{j\neq i}\left(2- (\psi^{(r)}_i-\psi^{(r)}_j)^2\right)\coth(\pi(\psi^{(r)}_i-\psi^{(r)}_j))\nonumber
\\
&&\qquad+\Bigg(\sum_j\Big[\sfrac14\left(\sfrac{1}{4}+(\psi^{(r)}_i\ms\psi^{(r\ps1)}_j\ms m)^2\right)\tanh(\pi(\psi^{(r)}_i\ms\psi^{(r\ps1)}_j\ms m))\nonumber\\
&&\qquad\qquad+\sfrac14\left(\sfrac{1}{4}+(\psi^{(r)}_i\ms\psi^{(r\ms1)}_j\ms m)^2\right)\tanh(\pi(\psi^{(r)}_i\ms\psi^{(r\ms1)}_j\ms m))\Big]\Big]\nonumber\\
&&\qquad\qquad\qquad +(m\to -m)\Bigg)\,.
\label{MIeom-quiver}
\end{eqnarray}
Equations (\ref{MIeom-quiver}) have a solution where $\psi^{(r)}_i=\psi^{(s)}_i$ for all $r$ and $s$.   Hence, if we take the same strong coupling ansatz as before,  we find  the eigenvalues set to $\psi^{(r)}_i=\phi_i$, where $\phi_i$ are the values in (\ref{MIsolution:strong}).
The free energy is $F_k=k F$, where $F$ is the free-energy in (\ref{MImatrix:free:energy}).  The Wilson loop is the same as in (\ref{MIwilson:loop:matrix}).

 \section{Supergravity comparisons}
\label{MIs-sugra}

In this section we compare the results for   strongly coupled 5d SYM with the corresponding results found using the AdS/CFT correspondence \cite{MIKallen:2012zn,MIMinahan:2013jwa}.    

We begin by reviewing the supergravity computation of the free   energy \cite{MIKallen:2012zn}.  We consider supergravity  on $AdS_7\times S^4$ where the $AdS_7$ boundary is $S^1\times S^5$.   The radii of $AdS_7$ and $S^4$ are $\ell$ and $\ell/2$ respectively, where $\ell=2\ell_{pl}(\pi N)^{1/3}$.  The $AdS_7$ metric in global coordinates is given by
\begin{equation}
ds^2=\ell^2(\cosh^2\rho\, d\tau^2+d\rho^2+\sinh^2\rho\, d\Omega_5^2)~,
\end{equation}
where $d\Omega_5^2$  is the unit 5-sphere metric.  The Euclidean time direction is compactified and has the identification $\tau\equiv \tau+2\pi R_6/r$, while $R_6$ and $r$ are the boundary radii of $S^1$ and $S^5$.

Under the AdS/CFT correspondence, the supergravity classical action equals the free   energy of the boundary field theory.  The action needs to be regulated by adding counterterms \cite{MIBalasubramanian:1999re,MIEmparan:1999pm,MIdeHaro:2000xn,MIAwad:2000aj}.  There can be scheme dependence in the regulation \cite{MIdeHaro:2000xn}, but we will follow a minimal subtraction type prescription, which is the normal procedure  when regulating the action.  The full action then has the form
\begin{equation}\label{MIIAdS}
I_{AdS}=I_{\rm{bulk}}+I_{\rm{surface}}+I_{\rm{ct}}\,,
\end{equation}
where
\begin{equation}
I_{\rm{bulk}}=-\frac{1}{16\pi G_N}\mbox{Vol}(S^4)\int d^7x \sqrt{g}\left(R-2\lambda\right)\
\end{equation}
 is the action in the bulk with Newton's constant related to the 11-dimensional Planck length as $G_N=16\pi^7\ell_{pl}^9$.  The other terms in (\ref{MIIAdS}) are  the surface contribution $I_{\rm{surf}}$ and the counterterm  $I_{\rm{ct}}$, written in terms of the boundary metric  and which cancels off divergences in $I_{\rm{bulk}}$.  One then finds the equations of motion
\begin{equation}
R-2\lambda=-\frac{12}{\ell^2}\,,
\end{equation}
and hence the action
 \begin{equation}\label{MIbulk}
I_{\rm{bulk}}=-\frac{1}{256\pi^8 \ell_{pl}^9}\left(\frac{\pi^2\ell^4}{6}\right)\frac{2\pi R_6}{r}\pi^3 (-12\ell^5)\int_0^{\rho_0}\cosh\rho\sinh^5\rho\, d\rho=\frac{4\pi R_6}{3\,r} N^3\sinh^6\rho_0\,.
\end{equation}
The integral diverges as $\rho_0\to\infty$  and corresponds to a UV divergence in the boundary theory.  We then set  $\lambda=e^{\rho_0}$ where $\lambda$ is the UV cutoff of the boundary theory, from which we obtain the expansion
\begin{equation}
\sinh^6\rho_0=\frac{1}{64}\lambda^{6}-\frac{3}{32}\lambda^{4}+\frac{15}{64}\lambda^{2}-\frac{5}{16}+{\rm O}(\lambda^{-2})\,.
\end{equation}
The surface term contributes to the divergent pieces, but not the finite part of (\ref{MIbulk}), while
the counterterm  cancels off the remaining divergent pieces with a minimal subtraction prescription.  Hence, the regularized action is \cite{MIEmparan:1999pm}
\begin{equation}
I_{AdS}=-\frac{5\pi R_6}{12\,r}N^3~.\label{MIsugra-final}
\end{equation}

The Wilson loop expectation value can also be computed using supergravity  \cite{MIBerenstein:1998ij}.  Here, one considers an M2 brane that wraps the Euclidean time direction and the equator, while the brane's third direction  drops straight down into the bulk.  The Wilson loop is then dominated by the extremum of the world-volume action of the M2 brane,
\begin{equation}
\langle W\rangle\sim e^{-T^{(2)}\int dV}\,,
\end{equation}
where the  tension of the brane is $T^{(2)}=\frac{1}{(2\pi)^{2}l_{p}^{3}}$. The M2 brane volume is given by
\begin{equation}
\int dV=l^{3}\int\limits_{0}^{\frac{2\pi R_{6}}{r}} d\tau
\int\limits_{0}^{2\pi}d\phi \int\limits_{0}^{\rho_{0}}d\rho\sinh(\rho)\cosh(\rho)
\end{equation}
Using the same UV cutoff as in (\ref{MIbulk}) we find
\begin{equation}
T^{(2)}\int dV=\frac{\pi N R_{6}}{r}\left(\lambda-2+\lambda^{-1}\right)\,.
\end{equation}
The integral is again regulated using the minimal subtraction procedure and gives the regulated Wilson loop
\begin{equation}
\langle W\rangle\sim \exp\left(\frac{2\pi N R_{6}}{r}\right)
\label{MIWilson:loop:sugra}
\end{equation}

Using the identification in (\ref{MIR6gym}) we see that (\ref{MIWilson:loop:sugra}) matches with (\ref{MIwilson:loop:matrix}) if $m=i/2$, which is the enhancement point to $\mathcal{N}=2$ supersymmetry \cite{MIKim:2012ava,MIMinahan:2015jta}.  However, for this value of $m$ the free-energy results in (\ref{MIsugra-final}) and (\ref{MImatrix:free:energy}) differ by a factor of 5/4.  Curiously, if we were to replace (\ref{MIR6gym}) with 
\begin{equation}\label{MIWLagree}
R_6=\frac{g_{YM}^2}{16\pi^2}\left(\frac{9}{4}+m^2\right)\,.
\end{equation}
such that the Wilson loops agree for any value of $m$, then the free-energies in  (\ref{MIsugra-final}) and (\ref{MImatrix:free:energy}) agree for $m=1/2$.   At present we do not have an explanation for this.
 
We can also compare supergravity results for the quiver.  This effectively replaces the $S^4$ with $S^4/\mathbb{Z}_k$, reducing the volume of this space by a factor of $k$, while at the same time replacing $N$ with $Nk$.  Therefore the free-energy becomes
\begin{equation}\label{MIIAdSk}
I_{AdS}=-\frac{5\pi R_6\,k^2}{12\,r}N^3~.
\end{equation}
Likewise, $R_6$ should be replaced with $R_6/k$.  Hence we have the same sort of matching/mismatching between the gauge theory and  supergravity  as in the unquivered theory.

\section{Discussion}
\label{MIsummary}

In this review we have shown how to extract perturbative results 
for five-dimensional $\mathcal{N}=1^*$ Yang-Mills theory on $S^5$.  There are of course many other interesting phenomena that one can explore.  For example, one can squash the $S^5$ \cite{MIQiu:2013pta} which modifies the determinant factors in (\ref{MImain-form-det}).  Interestingly, the free-energy is the same as in (\ref{MImatrix:free:energy}), multiplied by the ratio of the squashed to unsquashed volume factors \cite{MIQiu:2013pta}.  

Also of interest are phase transitions.  Here we will describe two types.  The first is a transition between a Yang-Mills phase and a Chern-Simons phase \cite{MIMinahan:2014hwa}.  Here it is possible to show that there is a third order phase transition when the ratio
\begin{equation}
\kappa\equiv \frac{8\pi^2}{k\,g_{YM}^2}
\end{equation}
equals the critical value
\begin{equation}
\kappa_c=\sqrt{\frac{27\pi\lambda}{2}}
\end{equation}
at weak coupling and
\begin{equation}
\kappa_c=4\pi^2(9+4m^2)\lambda
\end{equation}
at strong coupling.

Another type of phase transition occurs when taking the infinite volume limit \cite{MINedelin:2015mta},  in a manner similar to what happens in four dimensions \cite{MIRusso:2013qaa,MIRusso:2013kea}, there occur an infinite number of transitions as the actual 't Hooft coupling, $t\equiv g_{YM}^2N$, is taken to infinity. For an  adjoint hypermultiplet with mass $M$, at strong coupling one finds a series of critical points at \cite{MINedelin:2015mta}
\begin{equation}
t_c^{(n)}=\frac{8\pi^2}{M}(n+1)\,,\qquad n\in \mathbb{Z}_+\,,
\end{equation}
or in terms of the dimensionless 't Hooft coupling and mass parameter,
\begin{equation}
\lambda_c^{(n)}=\frac{8\pi^2}{m}(n+1)\,,\qquad n\in \mathbb{Z}_+\,.
\end{equation}

An important open problem is the mismatch of the free-energies between the 5d gauge theory calculation and the supergravity computation of the $(2,0)$ theory.  One possibility is that this is a scheme dependence problem.  In any event, we hope to see this discrepancy resolved in the near future. 
  
\bigskip\bigskip

\noindent{\bf\Large Acknowledgement}:
\bigskip

\noindent     This  research is supported in part by
Vetenskapsr\r{a}det under grant \#2012-3269.  I thank my co-authors J. K\"all\'en, A. Nedelin and M. Zabzine for very enjoyable collaborations.  I also thank the 
CTP at MIT   for kind
hospitality  during the course of this work.

\documentfinish

%% file: header.tex
\newcommand{\chapterauthor}[1]{
\begin{center}
{\bf \normalsize  #1}
\end{center}
}

\newcommand{\chapteraddress}[1]{
\begin{center}
{ \small \it \addressvalue}
\end{center}
}

\newcommand{\chapterabstract}[1]{
\vspace{\baselineskip}
\begin{center}
\textbf{\small Abstract}
\end{center}
#1}

\newcommand{\chapterheader}{

\chapter[\titlevalue{}  (by \shortauthorvalue)]{\titlevalue}
\label{Chapter\IDvalue}
\chapterauthor{\authorvalue}
\chapteraddress{\addressvalue}
\chapterabstract{\abstractvalue}
\tightmtctrue
\minitoc
}

\newcommand{\documentheader}{
\begin{flushright} \small
  \preprintvalue
 \end{flushright}

\begin{center}
{\bf \Large \titlevalue}
\end{center}

\chapterauthor{\authorvalue}
\chapteraddress{\addressvalue}
\chapterabstract{\abstractvalue}

\medskip

This is a contribution to the review volume ``Localization techniques
in quantum field theories'' (eds. V.~Pestun and M.~Zabzine) which
contains 17 Chapters available at \cite{ContributionSummary}

\tableofcontents
}

\newcommand{\ifvolume}[2]{\ifx\ifLONG\undefined#2\else#1\fi}

\newcommand{\documentfinish}{
\ifx\ifLONG\undefined
\bibliographystyle{bibreview} 
\bibliography{\IDvalue,review}  
\end{document}
\else
\addcontentsline{toc}{section}{References}
\input{\DIRvalue/\IDvalue.bbl}
\fi
}

\newcommand{\documentfinishBBL}{
\addcontentsline{toc}{section}{References}
\ifx\ifLONG\undefined
\input{\IDvalue.separate.bbl}
\end{document}
\else
\input{\DIRvalue/\IDvalue.volume.bbl}
\fi
}

\ifx\ifLONG\undefined
\def\volcite#1{Contribution \cite{Contribution#1}}
\else 
\def\volcite#1{Chapter \ref{Chapter#1}}
\fi

%% file: MI.bbl
\providecommand{\href}[2]{#2}\begingroup\raggedright\begin{thebibliography}{10}

\bibitem{ContributionSummary}
V.~Pestun and M.~Zabzine, eds., {\em Localization techniques in quantum field
  theory}, vol.~xx.
\newblock Journal of Physics A, 2016.
\newblock \href{http://arxiv.org/abs/1608.02952}{{\tt 1608.02952}}.
\newblock \url{https://arxiv.org/src/1608.02952/anc/LocQFT.pdf},
  \url{http://pestun.ihes.fr/pages/LocalizationReview/LocQFT.pdf}.

\bibitem{MIWitten:1995ex}
E.~Witten, ``{String Theory Dynamics in Various Dimensions},''
  \href{http://dx.doi.org/10.1016/0550-3213(95)00158-O}{{\em Nucl.Phys.} {\bf
  B443} (1995)  85--126},
\href{http://arxiv.org/abs/hep-th/9503124}{{\tt arXiv:hep-th/9503124
  [hep-th]}}.

\bibitem{ContributionKL}
S.~Kim and K.~Lee, ``Indices for 6 dimensional superconformal field theories,''
  {\em Journal of Physics A} {\bf xx} (2016)  000,
  \href{http://arxiv.org/abs/1608.02969}{{\tt 1608.02969}}.

\bibitem{MIDouglas:2010iu}
M.~R. Douglas, ``{On D=5 super Yang-Mills theory and (2,0) theory},''
  \href{http://dx.doi.org/10.1007/JHEP02(2011)011}{{\em JHEP} {\bf 1102} (2011)
   011},
\href{http://arxiv.org/abs/1012.2880}{{\tt arXiv:1012.2880 [hep-th]}}.

\bibitem{MILambert:2010iw}
N.~Lambert, C.~Papageorgakis, and M.~Schmidt-Sommerfeld, ``{M5-Branes,
  D4-Branes and Quantum 5D Super-Yang-Mills},''
  \href{http://dx.doi.org/10.1007/JHEP01(2011)083}{{\em JHEP} {\bf 1101} (2011)
   083},
\href{http://arxiv.org/abs/1012.2882}{{\tt arXiv:1012.2882 [hep-th]}}.

\bibitem{MIBolognesi:2011nh}
S.~Bolognesi and K.~Lee, ``{Instanton Partons in 5-dim SU(N) Gauge Theory},''
  \href{http://dx.doi.org/10.1103/PhysRevD.84.106001}{{\em Phys.Rev.} {\bf D84}
  (2011)  106001},
\href{http://arxiv.org/abs/1106.3664}{{\tt arXiv:1106.3664 [hep-th]}}.

\bibitem{MIBern:2012di}
Z.~Bern, J.~J. Carrasco, L.~J. Dixon, M.~R. Douglas, M.~von Hippel, {\em et
  al.}, ``{D = 5 Maximally Supersymmetric Yang-Mills Theory Diverges at Six
  Loops},''
\href{http://arxiv.org/abs/1210.7709}{{\tt arXiv:1210.7709 [hep-th]}}.

\bibitem{Papageorgakis:2014dma}
C.~Papageorgakis and A.~B. Royston, ``{Revisiting Soliton Contributions to
  Perturbative Amplitudes},''
  \href{http://dx.doi.org/10.1007/JHEP09(2014)128}{{\em JHEP} {\bf 09} (2014)
  128},
\href{http://arxiv.org/abs/1404.0016}{{\tt arXiv:1404.0016 [hep-th]}}.

\bibitem{MIKim:2012av}
H.-C. Kim and S.~Kim, ``{M5-branes from gauge theories on the 5-sphere},''
\href{http://arxiv.org/abs/1206.6339}{{\tt arXiv:1206.6339 [hep-th]}}.

\bibitem{MIKallen:2012zn}
J.~Kallen, J.~Minahan, A.~Nedelin, and M.~Zabzine, ``{$N^3$-behavior from 5D
  Yang-Mills theory},'' \href{http://dx.doi.org/10.1007/JHEP10(2012)184}{{\em
  JHEP} {\bf 1210} (2012)  184},
\href{http://arxiv.org/abs/1207.3763}{{\tt arXiv:1207.3763 [hep-th]}}.

\bibitem{MIKlebanov:1996un}
I.~R. Klebanov and A.~A. Tseytlin, ``{Entropy of Near Extremal Black
  P-Branes},'' \href{http://dx.doi.org/10.1016/0550-3213(96)00295-7}{{\em
  Nucl.Phys.} {\bf B475} (1996)  164--178},
\href{http://arxiv.org/abs/hep-th/9604089}{{\tt arXiv:hep-th/9604089
  [hep-th]}}.

\bibitem{MIHenningson:1998gx}
M.~Henningson and K.~Skenderis, ``{The Holographic Weyl Anomaly},'' {\em JHEP}
  {\bf 9807} (1998)  023,
\href{http://arxiv.org/abs/hep-th/9806087}{{\tt arXiv:hep-th/9806087
  [hep-th]}}.

\bibitem{MIMinahan:2013jwa}
J.~A. Minahan, A.~Nedelin, and M.~Zabzine, ``{5D Super Yang-Mills Theory and
  the Correspondence to AdS$_7$/CFT$_6$},''
  \href{http://dx.doi.org/10.1088/1751-8113/46/35/355401}{{\em J.Phys.} {\bf
  A46} (2013)  355401},
\href{http://arxiv.org/abs/1304.1016}{{\tt arXiv:1304.1016 [hep-th]}}.

\bibitem{MISeiberg:1996bd}
N.~Seiberg, ``{Five-Dimensional SUSY Field Theories, Nontrivial Fixed Points
  and String Dynamics},''
  \href{http://dx.doi.org/10.1016/S0370-2693(96)01215-4}{{\em Phys.Lett.} {\bf
  B388} (1996)  753--760},
\href{http://arxiv.org/abs/hep-th/9608111}{{\tt arXiv:hep-th/9608111
  [hep-th]}}.

\bibitem{MIIntriligator:1997pq}
K.~A. Intriligator, D.~R. Morrison, and N.~Seiberg, ``{Five-Dimensional
  Supersymmetric Gauge Theories and Degenerations of Calabi-Yau Spaces},''
  \href{http://dx.doi.org/10.1016/S0550-3213(97)00279-4}{{\em Nucl.Phys.} {\bf
  B497} (1997)  56--100},
\href{http://arxiv.org/abs/hep-th/9702198}{{\tt arXiv:hep-th/9702198
  [hep-th]}}.

\bibitem{MIKallen:2012va}
J.~Kallen, J.~Qiu, and M.~Zabzine, ``{The perturbative partition function of
  supersymmetric 5D Yang-Mills theory with matter on the five-sphere},''
  \href{http://dx.doi.org/10.1007/JHEP08(2012)157}{{\em JHEP} {\bf 1208} (2012)
   157},
\href{http://arxiv.org/abs/1206.6008}{{\tt arXiv:1206.6008 [hep-th]}}.

\bibitem{MIKim:2012ava}
H.-C. Kim and S.~Kim, ``{M5-branes from gauge theories on the 5-sphere},''
  \href{http://dx.doi.org/10.1007/JHEP05(2013)144}{{\em JHEP} {\bf 1305} (2013)
   144},
\href{http://arxiv.org/abs/1206.6339}{{\tt arXiv:1206.6339 [hep-th]}}.

\bibitem{ContributionQZ}
J.~Qiu and M.~Zabzine, ``Review of localization for 5D supersymmetric gauge
  theories,'' {\em Journal of Physics A} {\bf xx} (2016)  000,
  \href{http://arxiv.org/abs/1608.02966}{{\tt 1608.02966}}.

\bibitem{MIPestun:2007rz}
V.~Pestun, ``{Localization of Gauge Theory on a Four-Sphere and Supersymmetric
  Wilson Loops},'' \href{http://dx.doi.org/10.1007/s00220-012-1485-0}{{\em
  Commun.Math.Phys.} {\bf 313} (2012)  71--129},
\href{http://arxiv.org/abs/0712.2824}{{\tt arXiv:0712.2824 [hep-th]}}.

\bibitem{MIkurokawa}
S.-Y. Koyama and N.~Kurokawa, ``Zeta functions and normalized multiple sine
  functions,'' \href{http://dx.doi.org/10.2996/kmj/1134397767}{{\em Kodai Math.
  J.} {\bf 28} (2005) no.~3, 534--550}.
  \url{http://dx.doi.org/10.2996/kmj/1134397767}.

\bibitem{MINekrasov:1996cz}
N.~Nekrasov, ``{Five dimensional gauge theories and relativistic integrable
  systems},'' \href{http://dx.doi.org/10.1016/S0550-3213(98)00436-2}{{\em
  Nucl.Phys.} {\bf B531} (1998)  323--344},
\href{http://arxiv.org/abs/hep-th/9609219}{{\tt arXiv:hep-th/9609219
  [hep-th]}}.

\bibitem{MIFlacke:2003ac}
T.~Flacke, ``{Covariant quantization of N = 1, D = 5 supersymmetric Yang-Mills
  theories in 4D superfield formalism},''
\href{http://dx.doi.org/10.3204/DESY-THESIS-2003-047}{{\em
  DESY-thesis-2003-047} (2003)  }.

\bibitem{MIJafferis:2012iv}
D.~L. Jafferis and S.~S. Pufu, ``{Exact results for five-dimensional
  superconformal field theories with gravity duals},''
\href{http://arxiv.org/abs/1207.4359}{{\tt arXiv:1207.4359 [hep-th]}}.

\bibitem{MIJafferis:2010un}
D.~L. Jafferis, ``{The Exact Superconformal R-Symmetry Extremizes Z},'' {\em
  JHEP} {\bf 1205} (2012)  159,
\href{http://arxiv.org/abs/1012.3210}{{\tt arXiv:1012.3210 [hep-th]}}.

\bibitem{MIKallen:2012cs}
J.~Kallen and M.~Zabzine, ``{Twisted supersymmetric 5D Yang-Mills theory and
  contact geometry},'' \href{http://dx.doi.org/10.1007/JHEP05(2012)125}{{\em
  JHEP} {\bf 1205} (2012)  125},
\href{http://arxiv.org/abs/1202.1956}{{\tt arXiv:1202.1956 [hep-th]}}.

\bibitem{MIKim:2012qf}
H.-C. Kim, J.~Kim, and S.~Kim, ``{Instantons on the 5-Sphere and M5-Branes},''
\href{http://arxiv.org/abs/1211.0144}{{\tt arXiv:1211.0144 [hep-th]}}.

\bibitem{MIAssel:2012nf}
B.~Assel, J.~Estes, and M.~Yamazaki, ``{Wilson Loops in 5D ${\mathcal{N}}\!=1$
  SCFTs and AdS/CFT},''
\href{http://arxiv.org/abs/1212.1202}{{\tt arXiv:1212.1202 [hep-th]}}.

\bibitem{MIYoung:2011aa}
D.~Young, ``{Wilson Loops in Five-Dimensional Super-Yang-Mills},''
  \href{http://dx.doi.org/10.1007/JHEP02(2012)052}{{\em JHEP} {\bf 1202} (2012)
   052},
\href{http://arxiv.org/abs/1112.3309}{{\tt arXiv:1112.3309 [hep-th]}}.

\bibitem{MIMinahan:2015jta}
J.~A. Minahan and M.~Zabzine, ``{Gauge Theories with 16 Supersymmetries on
  Spheres},'' \href{http://dx.doi.org/10.1007/JHEP03(2015)155}{{\em JHEP} {\bf
  1503} (2015)  155},
\href{http://arxiv.org/abs/1502.07154}{{\tt arXiv:1502.07154 [hep-th]}}.

\bibitem{MIBalasubramanian:1999re}
V.~Balasubramanian and P.~Kraus, ``{A Stress Tensor for Anti-de~Sitter
  Gravity},'' \href{http://dx.doi.org/10.1007/s002200050764}{{\em
  Commun.Math.Phys.} {\bf 208} (1999)  413--428},
\href{http://arxiv.org/abs/hep-th/9902121}{{\tt arXiv:hep-th/9902121
  [hep-th]}}.

\bibitem{MIEmparan:1999pm}
R.~Emparan, C.~V. Johnson, and R.~C. Myers, ``{Surface Terms as Counterterms in
  the AdS / CFT Correspondence},''
  \href{http://dx.doi.org/10.1103/PhysRevD.60.104001}{{\em Phys.Rev.} {\bf D60}
  (1999)  104001},
\href{http://arxiv.org/abs/hep-th/9903238}{{\tt arXiv:hep-th/9903238
  [hep-th]}}.

\bibitem{MIdeHaro:2000xn}
S.~de~Haro, S.~N. Solodukhin, and K.~Skenderis, ``{Holographic Reconstruction
  of Space-Time and Renormalization in the AdS / CFT Correspondence},''
  \href{http://dx.doi.org/10.1007/s002200100381}{{\em Commun.Math.Phys.} {\bf
  217} (2001)  595--622},
\href{http://arxiv.org/abs/hep-th/0002230}{{\tt arXiv:hep-th/0002230
  [hep-th]}}.

\bibitem{MIAwad:2000aj}
A.~M. Awad and C.~V. Johnson, ``{Higher Dimensional Kerr - AdS Black Holes and
  the AdS / CFT Correspondence},''
  \href{http://dx.doi.org/10.1103/PhysRevD.63.124023}{{\em Phys.Rev.} {\bf D63}
  (2001)  124023},
\href{http://arxiv.org/abs/hep-th/0008211}{{\tt arXiv:hep-th/0008211
  [hep-th]}}.

\bibitem{MIBerenstein:1998ij}
D.~E. Berenstein, R.~Corrado, W.~Fischler, and J.~M. Maldacena, ``{The Operator
  product expansion for Wilson loops and surfaces in the large N limit},''
  \href{http://dx.doi.org/10.1103/PhysRevD.59.105023}{{\em Phys.Rev.} {\bf D59}
  (1999)  105023},
\href{http://arxiv.org/abs/hep-th/9809188}{{\tt arXiv:hep-th/9809188
  [hep-th]}}.

\bibitem{MIQiu:2013pta}
J.~Qiu and M.~Zabzine, ``{5D Super Yang-Mills on $Y^{p,q}$ Sasaki-Einstein
  Manifolds},'' \href{http://dx.doi.org/10.1007/s00220-014-2194-7}{{\em
  Commun.Math.Phys.} {\bf 333} (2015) no.~2, 861--904},
\href{http://arxiv.org/abs/1307.3149}{{\tt arXiv:1307.3149 [hep-th]}}.

\bibitem{MIMinahan:2014hwa}
J.~A. Minahan and A.~Nedelin, ``{Phases of Planar 5-Dimensional Supersymmetric
  Chern-Simons Theory},'' \href{http://dx.doi.org/10.1007/JHEP12(2014)049}{{\em
  JHEP} {\bf 1412} (2014)  049},
\href{http://arxiv.org/abs/1408.2767}{{\tt arXiv:1408.2767 [hep-th]}}.

\bibitem{MINedelin:2015mta}
A.~Nedelin, ``{Phase Transitions in 5D Super Yang-Mills Theory},''
\href{http://arxiv.org/abs/1502.07275}{{\tt arXiv:1502.07275 [hep-th]}}.

\bibitem{MIRusso:2013qaa}
J.~G. Russo and K.~Zarembo, ``{Evidence for Large-N Phase Transitions in
  ${\mathcal{N}}\!=2$* Theory},''
  \href{http://dx.doi.org/10.1007/JHEP04(2013)065}{{\em JHEP} {\bf 1304} (2013)
   065},
\href{http://arxiv.org/abs/1302.6968}{{\tt arXiv:1302.6968 [hep-th]}}.

\bibitem{MIRusso:2013kea}
J.~Russo and K.~Zarembo, ``{Massive ${\mathcal{N}}\!=2$ Gauge Theories at Large
  N},'' \href{http://dx.doi.org/10.1007/JHEP11(2013)130}{{\em JHEP} {\bf 1311}
  (2013)  130},
\href{http://arxiv.org/abs/1309.1004}{{\tt arXiv:1309.1004 [hep-th]}}.

\end{thebibliography}\endgroup

%% file: MIdef.tex
\usepackage{fullpage,epsfig,graphics,amsbsy,amssymb,cancel,slashed,mathrsfs}
\usepackage{psfrag,hyperref}
\usepackage{graphicx,color}
\usepackage{wrapfig}
\usepackage{amsmath}

\numberwithin{equation}{section}